\pdfoutput=1

\documentclass[11pt]{article}

\usepackage[final]{acl}
\usepackage{amsmath}
\usepackage{times}
\usepackage{latexsym}
\usepackage{array}
\usepackage{hyperref}
\usepackage[T1]{fontenc}

\usepackage[utf8]{inputenc}
\usepackage{multirow}
\usepackage{multicol}
\usepackage{booktabs}
\usepackage{microtype}

\usepackage{inconsolata}

\usepackage{graphicx}
\usepackage{amssymb}
\title{Beyond Single Labels: Improving Conversational Recommendation\\through LLM-Powered Data Augmentation}

\author{Haozhe Xu \thanks{\ \ These authors contributed equally.}, Xiaohua Wang\footnotemark[1], Changze Lv, Xiaoqing Zheng\thanks{\ \ Corresponding author} \\
         School of Computer Science, Fudan University, Shanghai, China \\
         \texttt{\{hzxu22,xiaohuawang22,czlv22\}@m.fudan.edu.cn}\\
         \texttt{zhengxq}@fudan.edu.cn}
\begin{document}
\maketitle
\begin{abstract}

Conversational recommender systems (CRSs) enhance recommendation quality by engaging users in multi-turn dialogues, capturing nuanced preferences through natural language interactions. 
However, these systems often face the false negative issue, where items that a user might like are incorrectly labeled as negative during training, leading to suboptimal recommendations.
Expanding the label set through data augmentation presents an intuitive solution but faces the challenge of balancing two key aspects: ensuring semantic relevance and preserving the collaborative information inherent in CRS datasets.
To address these issues, we propose a novel data augmentation framework that first leverages an LLM-based semantic retriever to identify diverse and semantically relevant items, which are then filtered by a relevance scorer to remove noisy candidates.
Building on this, we introduce a two-stage training strategy balancing semantic relevance and collaborative information.
Extensive experiments on two benchmark datasets and user simulators demonstrate significant and consistent performance improvements across various recommenders, highlighting the effectiveness of our approach in advancing CRS performance. \footnote{\ \ Code is available in \href{https://github.com/xu1110/FNSCRS}{https://github.com/xu1110/FNSCRS}}


\end{abstract}

\section{Introduction}

Conversational recommender systems (CRSs) provide recommendations by engaging users in multi-turn dialogues, capturing preferences through natural language interactions. This allows for explicit feedback and a nuanced understanding of user preferences \citep{gao2021advances, jannach2021survey}. 
As a result, CRSs are considered the next-generation recommender systems \citep{lin2023sparks, chen2024all}.

A CRS typically consists of a recommender responsible for producing recommendations and a dialogue generator tasked with generating the conversation.
While recent advances in large language models (LLMs) have alleviated the challenges associated with generating the conversation \citep{qin2024beyond}, significant challenges remain in optimizing the recommender.
\begin{figure}[t]
    \centering
    \includegraphics[scale=0.51]{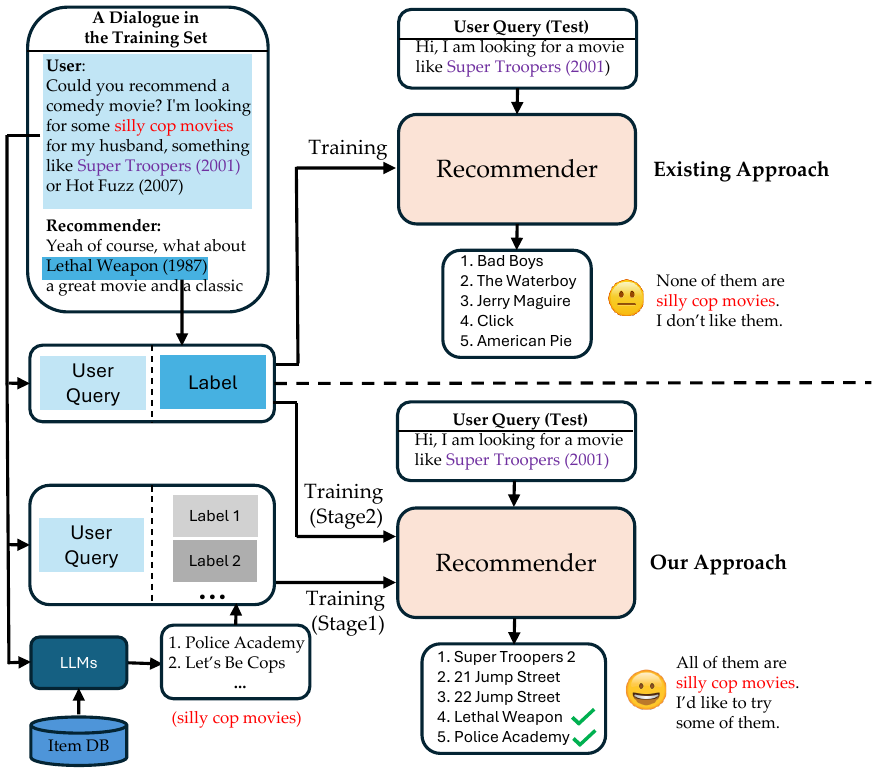}
    \caption{Comparison of existing and our proposed training processes. In the existing process, only one "silly cop movie" is labeled as positive, leading to unsatisfactory recommendations. Our approach considers multiple ``silly cop movies'' in the first training stage and uses the original label in the second training stage, which helps the recommender recommend ``silly cop movies'' while preserving collaborative information, such as the co-occurrence pattern of ``Super Troopers'' and ``Lethal Weapon''. The user finally accepted ``Lethal Weapon'' and ``Police Academy''.}
    \label{fig1}
\end{figure}
Figure \ref{fig1} illustrates a typical example of training a recommender, as commonly used in previous works.
In this example, when a user requests ``silly cop movies'', the recommender provides only one movie as a recommendation.
This user query and the recommended movie are treated as an input-label pair to train the recommender.
However, this approach does not account for other movies that the user might also be interested in, particularly other ``silly cop movies''.
Consequently, the recommender fails to suggest suitable movies, reducing the overall user experience. 
This issue is known as the ``false negative issue'' in traditional recommender systems \cite{ding2020simplify, liu2023ufnrec, wei2024llmrec}, where items that a user might like are incorrectly labeled as negative during training.
However, to the best of our knowledge, this issue has received limited attention in the context of CRS.


Expanding the label set is an intuitive solution to mitigate the false negative issue \citep{wei2024llmrec, liu2023ufnrec}. This approach incorporates additional items into the training process, enriching the label set. However, CRS datasets pose two unique challenges for data augmentation: ensuring semantic relevance and preserving collaborative information.

Firstly, CRS datasets are rich in semantic information \citep{he2023large}. Augmented labels should be diverse and semantically relevant to the dialogue context.
For example, if a user requests a ``silly cop movie'',  movies featuring a silly cop theme should be considered semantically relevant.
LLMs excel at understanding and generating semantically relevant content, making them well-suited for this task. By analyzing the dialogue context and item descriptions, LLMs can identify multiple semantically relevant items, effectively broadening the label set. However, relying solely on semantic relevance risks overlooking collaborative information, such as commonalities and trends among different users, which play an important role in effective recommendations \citep{zhu2024collaborative, zhang2023collm, he2024reindex}.

Secondly, collaborative information inherent in the original dataset should be preserved. While LLMs are adept at capturing semantic nuances, they often struggle to capture collaborative information effectively \citep{he2023large}. Over-reliance on LLM-suggested labels can lead to recommendations that prioritize semantic relevance at the expense of collaborative information, ultimately reducing user satisfaction.
Moreover, while collaborative information is important, it can sometimes introduce biases, such as popularity bias \citep{zhu2021popularity, zhao2022popularity}. Effectively utilizing collaborative information remains a challenging task.

To this end, we introduce a data augmentation approach to tackle the above challenges.
Firstly, we use an LLM-based semantic retriever to retrieve multiple and diverse potentially relevant items based on semantic similarity between the dialogue context and the item's descriptive text.
This step does not primarily consider collaborative information, thereby avoiding potential biases associated with it.
Then, we employ an LLM-based relevance scorer to assign fine-grained relevance scores to the retrieved items, filtering out those irrelevant ones. Items with high relevance scores are retained as augmented labels, forming the augmented dataset.
After that, we pre-train the recommender using the augmented dataset. This step enhances the recommender's ability to provide recommendations based on the semantic relevance between user preferences and items.
Then, we fine-tune the recommender on the original dataset to integrate the collaborative information. 
A label smoothing term can be added to control the integration of collaborative information.



We have conducted extensive experiments on two real-world CRS benchmarks, namely ReDial \citep{li2018towards} and INSPIRED \citep{hayati2020inspired}, as well as a recently proposed user simulator iEvaLM \citep{wang2023rethinking}.
Experimental results show that our approach can significantly and consistently improve the recommendation performance of different recommenders on different datasets.
These results demonstrated the effectiveness of our approach.

\section{Related Work}
\subsection{Conversational Recommender Systems}
Existing CRS can be categorized into two main types: attribute-based CRS and free-form CRS.
An attribute-based CRS explores user preferences by posing clarification questions about item attributes and then generating utterances using predefined templates \citep{sun2018conversational,zhang2018towards,lei2020estimation}.
A free-form CRS explores user preferences through various conversational modes, including chit-chat, question-answering, and recommendation attempts \citep{chen2019towards, zhou2020improving, yang2022improving,zhou2022c2, wang2022towards, zou2022improving, ren2022variational, wang2023rethinking, he2023large}.
In this paper, we focus on the free-form CRS due to its flexibility in engaging users in more natural and diverse interactions.

Developing an effective recommender is fundamental in constructing a CRS. Previous approaches include knowledge-graph-based \citep{chen2019towards, zhou2020improving, wang2022towards}, review-based \citep{lu2021revcore}, and text-based models \citep{yang2022improving, ravaut2024parameter}. Typically, these models use a limited number of positive items as training labels and treat others as negative, which can introduce noise. We propose a method to generate multiple labels for each dialogue.

\subsection{Synthetic Data in Conversational Recommendation}
To address the challenge of data scarcity, researchers have explored the use of synthetic data to train CRS. These efforts have primarily focused on generating \textbf{synthetic dialogues} through template-based dialogue creation \citep{zhao2023alleviating}, counterfactual dialogue simulation \citep{wang2023improving}, data-to-text dialogue generation \citep{lu2023august}, and LLM-based dialogue generation \citep{liang2024llm, wang2023improving2, kim-etal-2024-pearl}. In this paper, we investigate the generation of \textbf{synthetic labels} from existing dialogues—an underexplored yet complementary approach to existing works.


\subsection{Large Language Model-enhanced Conversational Recommendation}
With the development of large language models (LLMs), there are some early attempts to apply large language models for conversational recommendation.
These methods focus on dialogue management \citep{friedman2023leveraging, fang2024multi, feng2023large}, evaluation \citep{wang2023rethinking, huang2024concept}, dialogue generation \citep{qin2024beyond} or recommendation \citep{he2023large, he2024reindex, yang2024unleashing, xi2024memocrs}.
Although showing promising results, most of these methods suffer from high inference latency.
In this paper, we use LLMs to create synthetic data to support the training of downstream recommenders, which can avoid extensive inference latency.



\subsection{False Negative Issues in Traditional Recommender Systems}
The false negative issue occurs when a user is not exposed to certain items that they may be interested in, causing the recommender system to incorrectly assume that the user is not interested in them. These items are false negative samples (FNS).`

The current methods in traditional recommender systems can mainly be divided into two categories.
The first category treats FNS as noise and designs strategies to mitigate their negative impact. For instance, \citet{ding2020simplify} reduces the likelihood of FNS being selected during negative sampling, while \citet{zhang2024empowering} decreases the effect of noisy gradients from FNS through adversarial training.
The second category of methods involves augmenting the training dataset with potential FNS \citep{wei2024llmrec, liu2023ufnrec}.

Close to our work, \citet{wei2024llmrec} propose using a base recommender trained on the target datasets to retrieve potential FNS, and using an LLM to filter irrelevant ones in graph-based recommendations.
However, the key distinction in our approach is that we prioritize identifying semantically relevant items without initially considering collaborative information from the original dataset. In contrast, \citet{wei2024llmrec} utilize both collaborative and semantic information to construct their augmented dataset.
Our strategy offers two main advantages. Firstly, by initially ignoring collaborative information, we can retrieve a wider variety of items, reducing potential biases like popularity bias brought by collaborative information. This enhances the model's capability to learn the semantic connections between dialogue context and items. Secondly, our approach allows us to incorporate collaborative information at a later stage in a controlled manner, enabling a balanced integration of semantic relevance and collaborative information to suit various application scenarios.

\section{Method}
\begin{figure*}[t]
    \centering
    \includegraphics[scale=0.75]{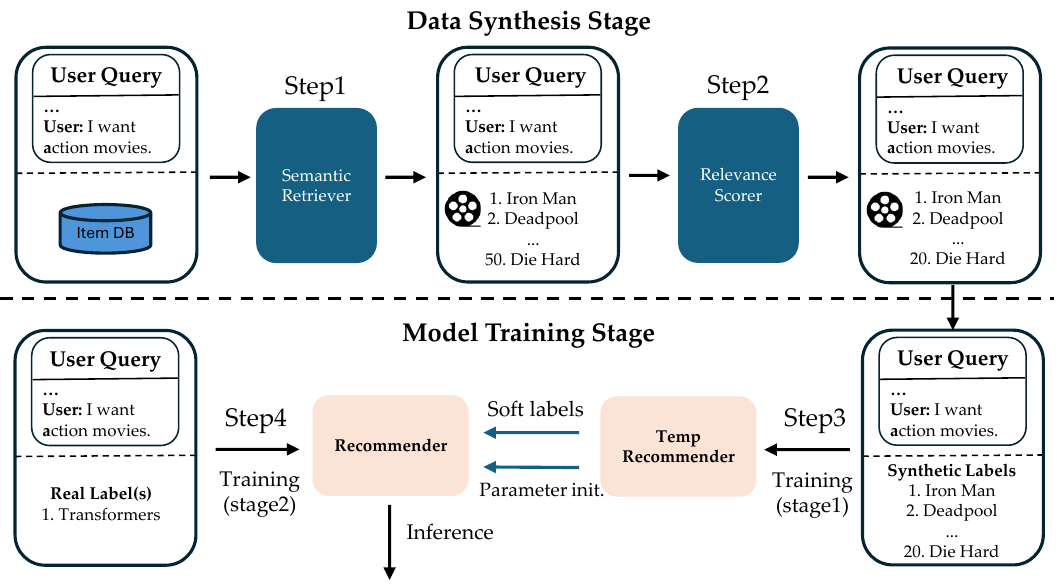}
    \caption{Overview of our approach. We first identify semantically relevant items as synthetic labels with LLMs-based semantic retriever and LLM-based relevance scorer. Then, we employ a two-stage training approach to pre-train the recommender with synthetic labels and fine-tune it with real labels.}
    \label{fig2}
\end{figure*}
Figure \ref{fig2} shows the overview of our approach. 
The main idea of our approach is to identify multiple semantically relevant items pre-dialogue (i.e. user query) to form the augmented dataset.
Then, a two-stage training scheme is employed to help the recommender focus on the semantic relationship between user preferences and items in the first training stage and integrate collaborative information in the second training stage.

\subsection{Data Synthesis Stage}
In practice, since the candidate pool of the item database can be very large, it is impractical for the LLM to score each candidate individually.
Therefore, we first use an LLM-based semantic retriever to retrieve a subset of potentially relevant items, thereby narrowing down the candidate set. Next, an LLM-based relevance scorer is employed to assign fine-grained relevance scores to these items.
\subsubsection{Relevant Items Retrieval}
Given item descriptive texts $\mathcal{I}=\lbrace{i_t}\rbrace^n_{t=1}$ and dialogue contexts $\mathcal{C}=\lbrace{c_t}\rbrace^m_{t=1}$, we first encode both of them into dense vector representations using an off-the-shelf LLM-based text encoder. 
Then, for each $c_t$ in $\mathcal{C}$, we retrieve the top-k most similar items $i_t$ through the max inner product search.
These items are treated as potential candidates for further filtering in the next step.

This retrieval process relies solely on semantic information, without considering collaborative information, thereby avoiding potential biases such as popularity bias introduced by collaborative information in the dataset.
This helps us cover a wide range of items instead of overly focusing on popular ones.
\subsubsection{Relevance Estimation} \label{sec:3.1.2}
After identifying potential relevant items, we use an LLM to perform detailed relevance estimations and filter out irrelevant samples.
To enhance efficiency and scalability, we employ a state-of-the-art LLM (e.g. GPT-4) to generate context-item-score triples using a chain-of-thought prompting approach \citep{wei2022chain}. These triples are then used to train a Gemma2-9b \citep{riviere2024gemma}, which assigns a relevance score ranging from 0 to 4 for each sample in the initial retrieval step. 
We retain context-item pairs with a relevance score exceeding a threshold value to construct the final synthetic training dataset $\mathcal{D}_{syn}=\lbrace{c_i, l_i}\rbrace^n_{i=1}$.
We choose 3.5 as the threshold value for its superior performance in our experiments. See Appendix \ref{apd:detail_relevance} for more details.

\subsection{Model Training Stage} \label{sec:train}
After we obtain the synthetic training dataset $\mathcal{D}_{syn}$, we can train the recommender with the synthetic training dataset $\mathcal{D}_{syn}$ and the original real-world training dataset $\mathcal{D}_{real}$.
A two-stage training scheme is designed to fully utilize the semantic information in the synthetic dataset while leveraging the collaborative information in the real-world dataset in a controllable manner.
\subsubsection{Pre-training the Recommender on the Synthetic Dataset}
In the first stage, the recommender is pre-trained using the synthetic dataset. This step aims to expose the model to a diverse range of semantically relevant items, helping it develop a foundational understanding of the semantic relationships between user preferences and the items. This is particularly important in conversational recommendations. By using synthetic data, this approach avoids introducing biases found in real-world data, such as specific popularity distributions.
A standard cross-entropy loss is applied for optimization:
\begin{equation}
    L_{pre} = -\sum_{i=1}^{N} \sum_{j=1}^{M} \ y_{i,j} \cdot \log P(i,j)
\end{equation} \label{eq1}
Where N, M denotes the number of dialogue contexts and items in the training dataset. 
$y_{i,j}$ denotes a binary label that is equal to 1 when a user-item pair is labeled as positive in the synthetic dataset.
$P(i,j)$ denotes a probability of the user-item pair being a positive pair calculated by the recommender.
The method of calculation varies among different recommenders.

\subsubsection{Fine-tuning the Recommender on the Real-world Dataset} 
After training with the synthetic dataset, the CRS can effectively provide recommendations based on semantic relevance between user preference and items.
However, relying solely on semantic relevance may not be sufficient to make effective recommendations.
In practice, collaborative information, referring to commonalities and trends in user behavior on target datasets/platforms, also plays an important role in making appropriate recommendations \citep{he2024reindex, zhang2023collm, zhu2024collaborative}.
This information is dataset/platform-dependent and can evolve due to semantic-independent factors (e.g. popularity trend), making it difficult for LLM to capture through semantic consistency \citep{he2023large}.
As a result, recommenders trained with LLM-suggested labels also struggle to consider collaborative information, leading to less appropriate and personalized recommendations.

To address this challenge, we propose to fine-tune the recommender on the original real-world dataset. 
This step ensures that the recommender can not only provide semantically relevant recommendations but also integrate collaborative information, enhancing its ability to reflect real-world user behaviors and preferences.
Similar to the pre-training stage, a cross-entropy loss is applied:
\begin{equation}
    L_{ce} = -\sum_{i=1}^{N} \sum_{j=1}^{M} \ y_{i,j} \cdot \log P(i,j)
\end{equation}
Here, $y_{i,j}$ denotes positive labels in the original real-world dataset.

Furthermore, a label smoothing term can be introduced during fine-tuning to balance the recommender's reliance on semantic relevance with collaborative information in making recommendations:
\begin{equation}
    L_{soft} = \sum_{i=1}^{N} D_{KL}(P(i), \hat{y_{i}})
\end{equation}
Where $\hat{y_{i,j}}$ is soft labels provided by the recommender before fine-tuning. $D_{KL}()$ denote the Kullback-Leibler Divergence.
Finally, we optimize the recommender through:
\begin{equation}
    L_{finetune} = L_{ce} + \alpha \cdot L_{soft}
\end{equation} \label{rec_loss}
Where a larger $\alpha$ indicates a decreased reliance on collaborative information, and vice versa.

\begin{table}[t]
    \centering
    \small
    \begin{tabular}{l|cccc}
        \hline
        Dataset & Train & Train Aug & Valid & Test\\
        \hline
        ReDial & 29810 & 377313 & 3318 & 4036 \\
        INSPIRED & 1404 & 15891 & 164 & 161 \\
        \hline
    \end{tabular}
    \caption{Statistics of the datasets after pre-processing. Numbers mean positive context-item pairs in the corresponding split. ``Train Aug'' denotes our synthetic dataset.}
    \label{tab:data_stat}
\end{table}

\begin{table*}[t]
    \centering
    \small
    \setlength{\tabcolsep}{16pt}
    \begin{tabular}{l|ccc|ccc}
    \hline
    \multirow{2}{*}{Models} & \multicolumn{3}{c|}{ReDial} & \multicolumn{3}{c}{INSPIRED} \\
    \cline{2-7}
    & R@1 & R@10 & R@50 & R@1 & R@10 & R@50 \\
    \hline
    \multicolumn{7}{c}{\textbf{Offline Dataset}} \\
    \hline
    BARCOR & 3.13 & 17.34 & 36.32 & 2.86 & 11.06 & 30.81 \\
    BARCOR + SD & 3.52 & 19.95 & 41.08 & 2.48 & 19.38 & 35.78 \\
    BARCOR + CFCRS & 3.50 & 18.98 & 38.46 & \textbf{4.72} & 20.50 & 35.16 \\
    BARCOR + ours & \textbf{4.31} & \textbf{21.26} & \textbf{43.84} & 3.73 & \textbf{21.12} & \textbf{43.11} \\
    \hline
    UniCRS & 3.53 & 19.60 & 40.50 & 3.97 & 20.00 & 40.66 \\
    UniCRS + SD & 3.08 & 20.27 & 42.26 & 4.64 & 20.53 & 38.94 \\
    UniCRS + CRCRS & \textbf{3.76} & 19.91 & 40.60 & \textbf{7.55} & 22.78 & \textbf{44.24} \\
    UniCRS + ours & \textbf{3.76} & \textbf{20.93} & \textbf{42.74} & 5.43 & \textbf{22.91} & 39.47 \\
    \hline
    Llama2 & 3.93 & 20.74 & 41.34 & 4.46 & 11.68 & 34.16 \\
    Llama2 + SD & 3.97 & 21.77 & 43.58 & 4.72 & 25.84 & 40.75 \\
    Llama2 + CRCRS & 4.27 & 21.57 & 42.24 & 5.22 & 19.35 & 37.39 \\
    Llama2 + ours & \textbf{4.46} & \textbf{22.37} & \textbf{44.20} & \textbf{9.32} & \textbf{28.26} & \textbf{50.93} \\
    \hline
    
    \hline
    \multicolumn{7}{c}{\textbf{User Simulator}} \\
    \hline
    BARCOR & 11.34 & 23.68 & 43.03 & 3.73 & 13.67 & 30.43 \\
    BARCOR + SD & 21.95 & 36.74 & 54.71 & 8.07 & 27.95 & 35.40 \\
    BARCOR + CFCRS & 19.87 & 34.17 & 51.23 & 14.91 & 24.22 & 37.27 \\
    BARCOR + ours & \textbf{27.77} & \textbf{44.69} & \textbf{63.35} & \textbf{24.84} & \textbf{40.37} & \textbf{61.49} \\
    \hline
    UniCRS & 20.46 & 36.44 & 55.72 & 8.07 & 23.60 & 40.37 \\
    UniCRS + SD & 22.07 & 39.12 & 59.78 & \textbf{15.53} & 24.84 & 40.99 \\
    UniCRS + CFCRS & \textbf{25.02} & 41.50 & 59.96 & 14.91 & 23.60 & 40.99 \\
    UniCRS + ours & 24.44 & \textbf{43.65} & \textbf{63.42} & 13.66 & \textbf{29.81} & \textbf{45.34} \\
    \hline
    Llama2 & 24.33 & 42.83 & 61.89 & 4.97 & 12.42 & 34.78 \\
    Llama2 + SD & 24.20 & 45.71 & 65.46 & 24.22 & 35.40 & 42.23 \\
    Llama2 + CFCRS & 27.95 & 45.19 & 63.08 & 26.09 & 39.75 & 51.55 \\
    Llama2 + ours & \textbf{28.65} & \textbf{50.00} & \textbf{71.42} & \textbf{34.78} & \textbf{57.76} & \textbf{73.29} \\
    \hline
    \end{tabular}
    \caption{Main experimental results. The best methods in each group are marked in bold. R@k denotes Recall@K. Standard deviations and significance tests can be found in Table \ref{tab:std}. }
    \label{tab:table1}
\end{table*}

\begin{table}[t]
    \centering
    \small
    \resizebox{\linewidth}{!}{\begin{tabular}{l|ccc|ccc}
    \hline
    \multirow{2}{*}{Models} & \multicolumn{3}{c|}{ReDial} & \multicolumn{3}{c}{INSPIRED} \\
    \cline{2-7}
    & R@1 & R@5 & R@10 & R@1 & R@5 & R@10 \\
    \hline
    BARCOR + ours & 4.31 & 14.00 & 21.26 & 3.73 & 15.40	& 21.12 \\
    Llama2 + ours & \textbf{4.46} & \textbf{14.52} & \textbf{22.37} & \textbf{9.32} & \textbf{20.19} & \textbf{28.26} \\
    \hline
    Zero-Shot(GPT3.5) & 3.57 & 11.54 & 16.33 & 4.35 & 11.80 & 17.39 \\
    Zero-Shot(GPT4o) & 4.05 & 11.62 & 17.20 & 6.21 & 16.15 & 21.74 \\
    \hline
    \end{tabular}}
    \caption{Comparisons with strong proprietary LLMs under zero-shot setting.}
    \label{tab:table_zero}
\end{table}

\section{Experiment}
To verify the effectiveness of our approach, we conduct experiments to figure out whether our approach can enhance the recommendation quality of different recommenders, and whether our approach can outperform other data augmentation technologies and other LLM-powered baselines.



\paragraph{Backbone models.}
We consider BARCOR \citep{wang2022barcor}, UniCRS \citep{wang2022towards} and Llama2-7b-chat-hf \citep{touvron2023llama} (denoted as Llama2) as backbone models.
These models are either the most representative CRS models or the most representative large language model.
See Appendix \ref{apd:imp_crs} for details.

\paragraph{Baselines.}
In addition to these backbone models, we introduce two additional data augmentation baselines. 

\begin{itemize}
\setlength{\itemsep}{0pt}
\setlength{\parsep}{0pt}
\setlength{\parskip}{0pt}
\item A false negative detection approach adapted from traditional recommender systems \cite{wei2024llmrec}, which considers both collaborative information and semantic information in the data synthesis stage by using supervised recommenders in the item retrieval stage. We denote this baseline as \textbf{self-distillation (SD)}. See Appendix \ref{apd:imp_fns} for details.
\item \textbf{CFCRS} \citep{wang2023improving}: It generates synthetic training dialogues through counterfactual data simulation.

\end{itemize}


\paragraph{Datasets.}

We conduct our experiments on two widely used datasets, namely ReDial proposed by \citet{li2018towards} and INSPIRED proposed by \citet{hayati2020inspired}, which focus on conversational movie recommendation.
We choose these two datasets because they are real-world datasets, which can lead to realistic evaluation.
The data statics are shown in Table \ref{tab:data_stat}. 

Previous work has found cases where items already mentioned in dialogue are also treated as labels, leading to the repetition of recommended items \citep{he2023large}. To ensure more reliable training and evaluation, we have removed these samples from the dataset and filtered out repeated items from the recommender's prediction list following \citet{he2023large}.

\paragraph{Evaluation Setups.}
We conduct two types of evaluations, namely evaluation with the offline dataset and evaluation with the user simulator.
The former is the most widely used evaluation protocol in previous work \cite{wang2022towards, wang2022barcor}. The latter is a newly proposed protocol that uses an LLM to interact with the CRS, which mimics real-world human behavior.
We use iEvaLM \citep{wang2023rethinking} as the user simulation protocol.
Following previous work, we apply Recall@k k=\{1,10,50\} as evaluation metrics.

\paragraph{Implementation Details.}
In the retrieval step, we retrieve the 50 most similar items for each dialogue.
We use GritLM \citep{muennighoff2024generative} as the text encoder due to its superior text representation ability.
We follow \citet{wang2023improving} to perform the data processing.
Since we focus on recommendation task, we implement dialogue generation by filling recommended items into templates like \citet{wang2022towards}. We will discuss other variants in Section \ref{sec:dis} and Appendix \ref{apd:case_study_generation}.
We run each experiment with the offline dataset evaluation five times and report the average numbers. 
Due to the extensive cost, we run each experiment with the user simulator once.

\subsection{Main Results}
Table \ref{tab:table1} shows the results on two benchmark datasets, including both evaluation setups.

Compared with base models, our approach can consistently and significantly improve the performance of all three representative CRS models under both settings.
These results demonstrated the effectiveness of our proposed approach. We will conduct further analysis on how these improvements are achieved in Section \ref{sec: data} and Section \ref{sec:ana_cf}.

Compared with self-distillation, our approach yields greater improvements, confirming the effectiveness of disregarding collaborative information during data synthesis while integrating it through two-stage training. 
Further analysis of this will be provided in Section \ref{sec:ana_fns}.



Compared to CFCRS, our approach demonstrates superior performance in most cases. This improvement can be attributed to our method’s utilization of the LLM’s semantic understanding ability to augment training labels based on existing dialogues, thereby ensuring semantic consistency between dialogues and labels. In contrast, CFCRS generates new dialogues through entity-based counterfactual data simulation, which lacks deep semantic understanding and may result in lower-quality dialogues. Additionally, our approach remains compatible with CFCRS, as the latter’s generated dialogues still suffer from the false negative issue. Exploring the potential synergy between generating new dialogues (as in CFCRS) and augmenting labels (as in our method) is an interesting direction for future work.

\subsection{Comparison with Zero-shot LLMs}
To further verify the effectiveness of our approach, we compare our approach with proprietary LLMs in the zero-shot setting on offline datasets \citep{he2023large}.
As shown in Table \ref{tab:table_zero}, our approach achieves superior performance compared to strong proprietary LLMs. Notably, our method outperforms these proprietary models despite using significantly smaller models. This result highlights the importance of sufficient training in adapting LLMs for conversational recommendation tasks.

\section{In-depth Analysis} \label{sec:ana}
In this section, we conduct an in-depth analysis of how our approach improves the overall performance of recommenders and the rationale behind our approach. We investigate how augmented data improve the performance of recommenders in Section \ref{sec: data}, how our two-stage training scheme helps recommenders capture collaborative information in Section \ref{sec:ana_cf}, how our approach outperforms the self-distillation baseline in Section \ref{sec:ana_fns}, and how the quantity and quality of synthetic data affect the performance of recommenders in Section \ref{sec:ana4}.





\subsection{Effect of Augmented Data} \label{sec: data}
\paragraph{Our approach can effectively retrieve multiple semantically relevant items based on user preferences, making the recommender's recommendations more semantically relevant, and alleviates the false negative issue.}
\begin{table}[t]
    \centering
    \small
    \setlength{\tabcolsep}{8pt}
    \begin{tabular}{l|cc|cc}
    \hline
    \multirow{2}{*}{Models} & \multicolumn{2}{c|}{ReDial} & \multicolumn{2}{c}{INSPIRED} \\
    \cline{2-5}
    & S@1 & S@5 & S@1 & S@5 \\
    \hline
    BARCOR & 2.94 & 2.79 & 1.05 & 1.26 \\
    BARCOR + ours & \textbf{3.46} & \textbf{3.11} & \textbf{2.79} & \textbf{2.67}\\
    \hline
    UniCRS & 3.04 & 2.80 & 1.82 & 1.90 \\
    UniCRS + ours & \textbf{3.29} & \textbf{3.04} & \textbf{2.20} & \textbf{2.15} \\
    \hline
    Llama2 & 3.21 & 2.97 & 0.79 & 1.23 \\
    Llama2 + ours & \textbf{3.34} & \textbf{3.08} & \textbf{3.17} & \textbf{2.85} \\
    \hline

    \end{tabular}
    \caption{Results on semantic relevance. S@k means the average relevance score of top-k recommendations.}
    \label{tab:table_rel}
\end{table}

To evaluate the CRS's ability to provide semantically relevant items, we conduct experiments to assess the semantic relevance between the dialogue context and the recommended items. Inspired by the effectiveness of LLMs as automatic evaluators for conversational recommendations \cite{wang2023rethinking, huang2024concept}, we use GPT-4 to assign relevance scores (0-4) to the top-k recommended items based on the dialogue context.
As shown in Table \ref{tab:table_rel}, our approach consistently improves the recommenders' capacity to recommend items that are semantically relevant to the dialogue context, which is important for effective recommendations. To verify the reliability of GPT-4's assessments, we sample 625 dialogue-item pairs for human evaluation. The Pearson correlation between humans and GPT-4 is 0.7, indicating high agreement. Moreover, some of these augmented items are likely to be false negative samples, alleviating the false negative issue.

Several cases can be found in Appendix \ref{apd:case_study} to illustrate how augmented items affect the performance of the recommender.

\subsection{Effect of Training Process} \label{sec:ana_cf}
\begin{table}[t]
\centering
\small
\resizebox{\linewidth}{!}{\begin{tabular}{l|cccc}
\hline
Models & R@10 & R@50 & CS@10 & CS@50\\
\hline
BARCOR & 17.34 & 36.32 & 3.49 & 2.36 \\
BARCOR + ours & \textbf{21.26} & \textbf{43.84} & \textbf{3.59} & \textbf{2.41} \\
\hline
w/o FT & 12.55 & 32.27 & 1.80 & 1.38 \\
mix & 13.1 & 34.4 & 1.87 & 1.52 \\
\hline
\hline
UniCRS & 19.6 & 40.5 & 3.49 & 2.34 \\
UniCRS+ours & \textbf{20.93} & \textbf{42.74} & \textbf{3.58} & \textbf{2.38} \\
\hline
w/o FT & 12.09 & 30.94 & 1.73 & 1.36 \\
mix & 13.2 & 34.9 & 1.89 & 1.54 \\
\hline
\hline
Llama2 & 20.74 & 41.34 & \textbf{3.60} & \textbf{2.48} \\
Llama2+ours & \textbf{22.37} & \textbf{44.20} & 3.53 & 2.44 \\
\hline
w/o FT & 11.52 & 31.10 & 1.62 & 1.29 \\
mix & 12.49 & 33.45 & 1.76 & 1.42\\
\hline
\end{tabular}}
\caption{Comparison of different training schemes on the ReDial dataset. ``w/o FT'' denotes removing the fine-tuning stage on the original dataset, ``mix'' denotes directly mixing augmented labels and original labels to train the recommender. ``CS@k'' denotes the average collaborative score of the top-k recommendation.}
\label{tab:table4}
\vspace{-3mm}
\end{table}

\paragraph{Our two-stage training process helps the recommender to capture collaborative information, resulting in more appropriate recommendations.}
To verify the importance of collaborative information in generating appropriate recommendations and to assess the effectiveness of our training scheme, we introduce two variants on ReDial dataset:
One is to remove the fine-tuning stage, and the other is to directly mix both synthetic data and real data to train the recommender.
To quantize how much collaborative information the recommender can capture, we first train a collaborative filtering model (e.g. SASRec \citep{kang2018self}) on the original dataset to model item co-occurrence patterns, which is one of the most important collaborative information \cite{sarwar2001item, he2023large}.
After that, we use the model to generate collaborative scores (denoted as CS) by calculating the co-occurrence probability (unnormalized logits) between the items recommended by the recommender and the items mentioned in the dialogue context, which assess the recommender's ability to capture item co-occurrence pattern.
As shown in Table \ref{tab:table4}, training only with the synthetic dataset or mixing the two datasets results in low collaborative scores, leading to lower overall performance compared to our approach.
This finding verified the importance of collaborative information in making appropriate recommendations and demonstrates the effectiveness of our approach in integrating collaborative information.
\begin{figure}[t]
    \centering
    \includegraphics[scale=0.125]{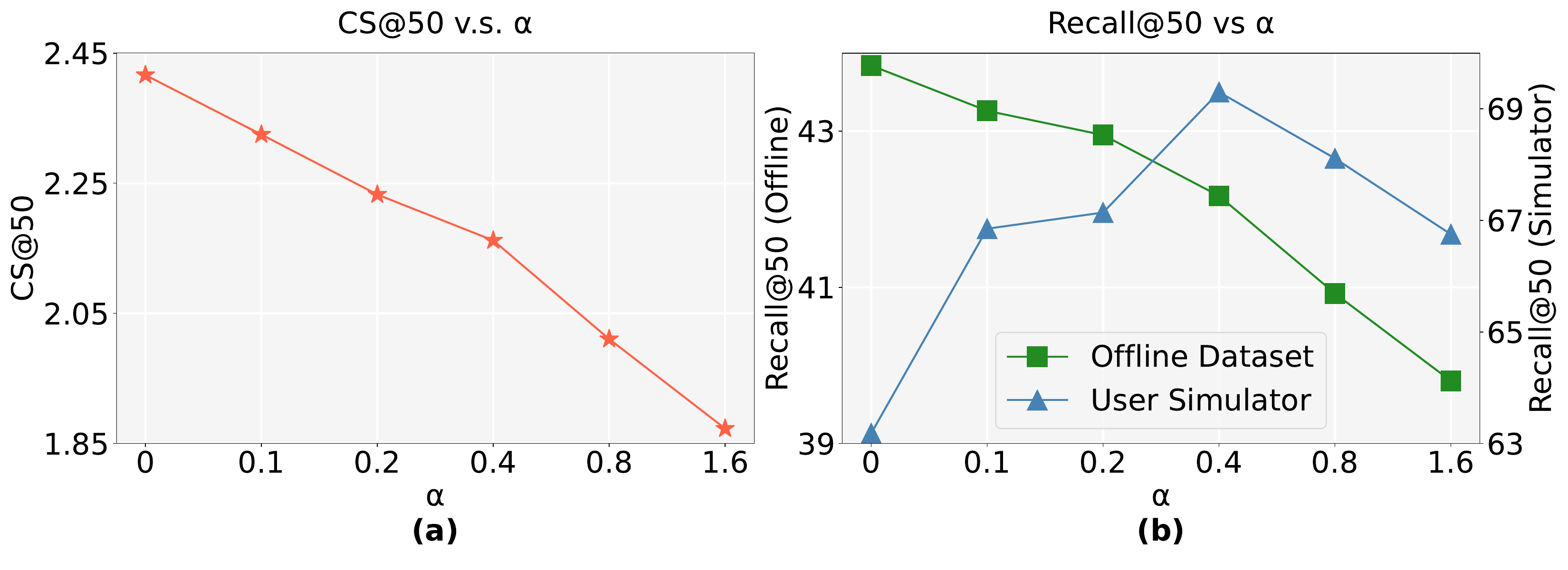}
    \caption{CS@50 and the Recall@50 v.s. different value of $\alpha$.}
    \label{fig3}
\end{figure}

\paragraph{Including soft labels can control the integration of collaborative information, offering flexible control to adapt to different application scenarios}
To illustrate the impact of soft labels as described in Equation \ref{eq1}, we conduct additional experiments using both the offline dataset and the user simulator settings on the ReDial dataset.
We use BARCOR trained with different $\alpha$ values in Equation \ref{rec_loss} to illustrate how soft labels affect the recommender.
Figure \ref{fig3} (a) indicates that increasing $\alpha$ decreases the ability of the recommender to capture collaborative information, enabling it to focus more on semantic relevance while providing recommendations.
This verifies the effectiveness of our approach to balancing semantic relevance and collaborative information through soft labels.

Figure \ref{fig3} (b) shows that when $\alpha$ is less than 0.4, increasing $\alpha$ positively impacts the performance in the user simulator but negatively impacts the performance in the offline dataset.
This difference arises because the user simulator tends to mention fewer items and provide more detailed textual preferences \citep{wang2023rethinking}, making semantic relevance more important than collaborative information when making recommendations. 
In contrast, users in the offline dataset tend to mention more items and offer less detailed textual preferences, so collaborative information is more important when making recommendations.
An example can be found in Appendix \ref{apd:example} to illustrate the difference between the offline dataset and the user simulator.
However, when $\alpha$ is larger than 0.4, continually increasing $\alpha$ can decrease the performance in both evaluations, which indicates the importance of collaborative information even when detailed textual preference is provided.
Our findings suggest that the importance of collaborative information may vary across different scenarios, emphasizing the value of our approach to offer control over the integration of collaborative information.



\subsection{Comparison with Self-distillation} \label{sec:ana_fns}
\paragraph{Disregarding collaborative information and focusing on semantic relevance in the item retrieval stage help us cover a wider range of items, leading to better long-tail performance.}
As mentioned previously, the key difference between our approach and self-distillation lies in the item retrieval stage. In our approach, we use a semantic retriever and do not incorporate collaborative information from the dataset, while self-distillation uses the base recommender as the retriever, which takes collaborative information into account.
To figure out how our strategy leads to better overall performance, we conduct an additional analysis on the ReDial dataset.
Table \ref{tab:fns} presents the number of relevant items retrieved by different retrievers, the number of distinct items covered by each retriever, and the long-tail performance for models using different retrievers (we denote items mentioned fewer than 4 times as long-tail items following \citet{zhao2023alleviating}).
While using the base recommender as a retriever can retrieve a similar number of relevant items, it covers significantly fewer distinct items compared to our semantic retriever. This is due to the popularity bias introduced by collaborative information, which causes the retriever to over-prioritize popular items while neglecting less popular ones.
As a result, our approach shows better long-tail performance, which leads to better overall recommendation performance.

\begin{table}[t]
    \centering
    \small
    \resizebox{\linewidth}{!}{\begin{tabular}{l|cccc}
    \hline
    Model & Count & Coverage & TR@10 & TR@50\\
    \hline
    BARCOR+SD & 368815 & 51.4\% & 1.24 & 3.52 \\
    BARCOR+ours & 377313 & 98.1\% & \textbf{2.14} & \textbf{8.7}\\
    \hline
    UniCRS+SD & 385567 & 57.8\% & 1.90 & 6.03 \\
    UniCRS+ours & 377313 & 98.1\% & \textbf{2.71} & \textbf{9.77} \\
    \hline
    Llama2+SD & 415950 & 71.7\% & 2.89 & 7.82 \\
    Llama2+ours & 377313 & 98.1\% & \textbf{3.01} & \textbf{9.37} \\
    \hline
    \end{tabular}}
    \caption{Comparison with self-distillation. ``Count'' and ``Coverage'' indicate how many relevant items are retrieved by different retrievers and how many distinct items can different retrievers cover. ``TR@k'' denotes Tail\_Recall@k, which computes the recall rate only for long-tail items.}
    \label{tab:fns}
\end{table}

\subsection{Effect of Data Quantity and Quality} \label{sec:ana4}
\paragraph{Including more synthetic data can lead to better performance.}
\begin{figure}[t]
    \centering
    \includegraphics[scale=0.12]{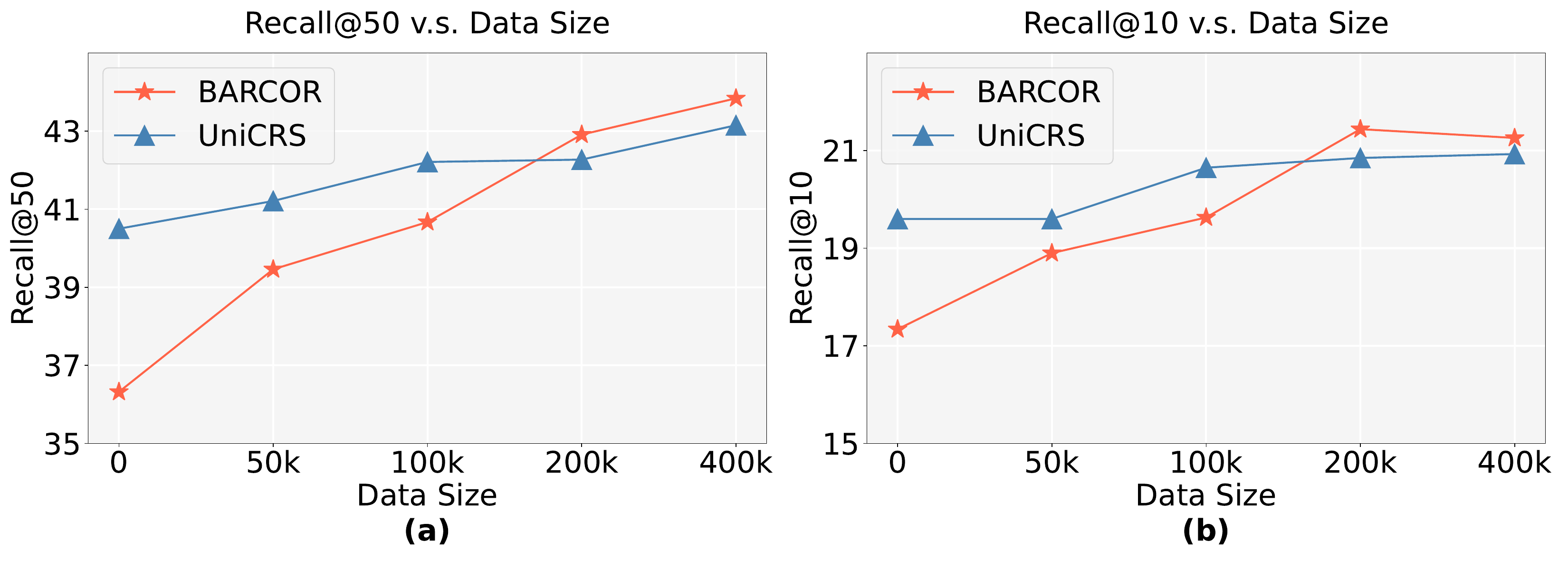}
    \caption{Performance v.s. different amounts of synthetic data on the ReDial dataset. ``Data Size'' denotes the number of synthetic data used during pre-training.}
    \label{fig4}
\end{figure}
As shown in figure \ref{fig4}, as the amounts of synthetic data increase, the performance of the recommender also increases, and the performance does not seem to converge.
This finding supports our assumption that incorporating a variety of labels can help boost the recommender's recommendation performance.
Meanwhile, the training time is directly proportional to the number of training samples. 
Therefore, in practice, others may need to balance the trade-off between training time and the performance of the CRS.

\begin{table}[t]
    \centering
    \small
    \begin{tabular}{l|ccc}
    \hline
    Model & R@1 & R@10 & R@50\\
    \hline
    BARCOR+ours & \textbf{4.31} & \textbf{21.26} & \textbf{43.84} \\
    \hline
    w/o scorer & 3.92 & 21.21 & 43.29\\
    \hline
    \hline
    UniCRS+ours & \textbf{3.76} & \textbf{20.93} & \textbf{42.74} \\
    \hline
    w/o scorer & 3.37 & 18.26 & 41.39 \\
    \hline
    \end{tabular}
    \caption{Ablation study on the LLM relevance scorer on ReDial dataset. ``w/o'' scorer denotes removing the LLM relevance scorer.}
    \label{tab:llm_score}
\end{table}
\paragraph{Without the LLM relevance scorer, the performance drops due to the noisy training data.}
As shown in table \ref{tab:llm_score}, if we remove the LLM filter, the performance drops consistently.
This finding verifies the importance of fine-grained relevance estimation with the LLM relevance scorer, which can effectively filter irrelevant items.


\section{Discussions on Dialogue Generation} \label{sec:dis}
Most CRS include a recommendation module to provide recommendations and a conversation module to generate response for the user.
Unlike previous work that treats these two modules as a unified system \cite{wang2022barcor, wang2022towards}, we follow the modular design principle suggested by \citet{zhang2024recwizard} and optimize the recommendation module independently, enabling the recommendation module to flexibly collaborate with various generation modules, such as filling recommended items into template utterances or prompting LLMs to generate responses based on the recommendations.
Although our approach focuses solely on recommendation quality, we still offer several suggestions and observations related to dialogue generation: (1) The modular design can potentially enhance dialogue generation quality by allowing the recommendation module to be combined with more advanced generation modules, rather than treating both modules as a unified system. (2) Improved recommendation quality can benefit dialogue generation and making the generated responses more coherent. For further illustrations, see Appendix \ref{apd:case_study_generation}.

\section{Conclusion}

We present a data augmentation approach to augment multiple labels for each dialogue in the CRS training dataset.
We first identify multiple semantically relevant items with an LLM-based semantic retriever and relevance scorer for each dialogue, which serve as augmented labels.
Then, we design a two-stage training scheme to first pre-train the recommenders with augmented data to capture semantic information, and then fine-tune them with real-world data to integrate collaborative information in a controllable manner.
Experimental results have shown a significant improvement in various settings, highlighting our approach's effectiveness.



\section*{Limitations}

Though our approach shows promising results.
However, we acknowledge two limitations of our approach.
Firstly, we only focus on the movie domain, which is the most commonly used domain in conversational recommendation. In the future, if widely used datasets in other domains are available, we should adapt our approach to other domains to verify the generalization ability of our approach.
Secondly, including additional training samples can add up to additional training costs. We leave it to future work to develop more cost-efficient training technology.

\bibliography{custom}

\clearpage

\appendix


\section{Details of the Relevance Estimation Step} \label{apd:detail_relevance}
As discussed in Section \ref{sec:3.1.2}, we employ an LLM-based relevance scorer to filter noisy candidates. First, we sample five items pre-dialogue and use GPT-4 (GPT-4o-mini) to assign each a relevance score (0–4) based on its alignment with the dialogue. The scoring prompt is provided in Table \ref{tab:prompt_llmreceval}. These scored samples are then used to train a Gemma2-9b model for efficient inference, optimized via MSE loss.

\begin{equation}
\mathcal{L}_{ce} = \frac{1}{N} \sum_{i=1}^{N} \left(  (f_\theta(c_i, l_i) - s_i)^2  \right)
\end{equation}

Where $f_\theta(c_i, l_i)$ denotes the score assigned by the Gemma2-9b model to the dialogue-product pair.  
Specifically, the computation of $f_\theta(c_i, l_i)$ is performed as follows: the dialogue $c_i$ and item description $l_i$ are concatenated into the prompt sequence $t_{1:n}$, where $n$ denotes the length of the prompt. This sequence is then fed into the Gemma2-9b model. The last hidden state $h_n$ from the final hidden layer of the model is extracted and passed through a trainable fully connected layer $W$, which maps the 4096-dimensional $h_n$ to a 1-dimensional score output $f_\theta(c_i, l_i)$:  

\begin{equation}  
f_\theta(c_i, l_i) = W(h_n)  
\end{equation}

The Gemma2-9B model is fine-tuned using LoRA \citep{hulora} with a rank of 8.

\section{Standard Deviations}
We report the mean, the standard deviations, and the significance test in our main experiments with offline dataset evaluation in Table \ref{tab:std}.
\begin{table*}[t]
    \centering
    \small
    \begin{tabular}{l|ccc|ccc}
    \hline
    \multirow{2}{*}{Models} & \multicolumn{3}{c|}{ReDial} & \multicolumn{3}{c}{INSPIRED} \\
    \cline{2-7}
    & R@1 & R@10 & R@50 & R@1 & R@10 & R@50 \\
    \hline
    \multicolumn{7}{c}{\textbf{Offline Dataset}} \\
    \hline
    BARCOR & 3.13 (0.27) & 17.34 (0.29) & 36.32 (0.21) & 2.86 (1.29) & 11.06 (1.83) & 30.81 (2.83) \\
    BARCOR + SD & 3.52 (0.31) & 19.95 (0.26) & 41.08 (0.63) & 2.48 (1.13) & 19.38 (1.29) & 35.78 (2.87) \\
    BARCOR + CFCRS & 3.50 (0.11) & 18.98 (0.33) & 38.46 (0.46) & \textbf{4.72 (0.56)} & 20.50 (1.96) & 35.16 (1.62) \\
    BARCOR + ours & \textbf{4.31$^{*\dagger}$ (0.21)} & \textbf{21.26$^{*\dagger}$ (0.49)} & \textbf{43.84$^{*\dagger}$ (0.26)} & 3.73 (1.52) & \textbf{21.12$^{*}$ (0.98)} & \textbf{43.11$^{*\dagger}$ (1.68)} \\
    \hline
    UniCRS & 3.53 (0.12) & 19.60 (0.31) & 40.50 (0.25) & 3.97 (1.05) & 20.00 (1.18) & \textbf{40.66 (1.37)} \\
    UniCRS + SD & 3.08 (0.17) & 20.27 (0.39) & 42.26 (0.51) & 4.64 (0.66) & 20.53 (0.66) & 38.94 (0.98) \\
    UniCRS + CRCRS & \textbf{3.76 (0.23)} & 19.91 (0.22) & 40.60 (0.35) & \textbf{7.55 (0.76)} & 22.78 (1.79) & \textbf{44.24 (1.44)} \\
    UniCRS + ours & \textbf{3.76$^{\dagger}$ (0.21)} & \textbf{20.93$^{*}$ (0.59)} & \textbf{42.74$^{*}$ (0.39)} & 5.43 (1.36) & \textbf{22.91$^{*\dagger}$ (1.37)} & 39.47 (0.36) \\
    \hline
    Llama2 & 3.93 (0.35) & 20.74 (0.53) & 41.34 (0.79) & 4.46 (1.21) & 11.68 (1.02) & 34.16 (1.58) \\
    Llama2 + SD & 3.97 (0.22) & 21.77 (0.40) & 43.58 (0.43) & 4.72 (0.94) & 25.84 (1.13) & 40.75 (1.84) \\
    Llama2 + CRCRS & 4.27 (0.29) & 21.57 (0.34) & 42.24 (0.53) & 5.22 (1.29) & 19.35 (1.24) & 37.39 (1.11) \\
    Llama2 + ours & \textbf{4.46$^{*\dagger}$ (0.30)} & \textbf{22.37$^{*\dagger}$ (0.31)} & \textbf{44.20$^{*}$ (0.73)} & \textbf{9.32$^{*\dagger}$ (0.88)} & \textbf{28.26$^{*}$ (2.18)} & \textbf{50.93$^{*\dagger}$ (1.13)} \\
    \hline

    \end{tabular}
    \caption{Mean and standard deviations (stds.) of main experimental results in offline dataset evaluation. The numbers outside the parentheses are the mean and the numbers inside the parentheses are stds. $*$denotes significant improvement compared with the base model, and $\dagger$ denotes significant improvement compared with self-distillation. (p<0.05)}
    \label{tab:std}
\end{table*}

\section{Introduction and Implementation Details of Baselines} \label{apd:detail_baseline}

\begin{table}[t]
    \centering
    \small
    \begin{tabular}{l|l|ccc}
    \hline
    Model & $\omega_3$ & R@1 & R@10 & R@50\\
    \hline
    \multicolumn{5}{c}{\textbf{Original One-stage Training}} \\
    \hline
    BARCOR & 0.2 & 2.85 & 16.38 & 37.41 \\
    BARCOR & 0.4 & 3.07 & 17.78 & 38.45 \\
    BARCOR & 0.6 & 3.07 & 17.66 & 38.33 \\
    BARCOR & 0.8 & 2.70 & 18.14 & 39.12 \\
    BARCOR & 1 & 2.80 & 18.38 & 39.74 \\
    BARCOR & 2 & 2.87 & 17.89 & 40.14 \\
    BARCOR & 5 & 2.45 & 17.98 & 40.70 \\
    BARCOR & 10 & 2.40 & 17.91 & \textbf{41.10} \\
    \hline
    \multicolumn{5}{c}{\textbf{Our Two-stage Training}} \\
    \hline
    BARCOR & - & \textbf{3.52} & \textbf{19.95} & 41.08 \\
    \hline
    \end{tabular}
    \caption{Performance comparison of our two-stage training scheme and the original one-stage training scheme in implementing the self-distillation baseline with BARCOR on the ReDial dataset. $\omega_3$ denotes the proportion of synthetic data per batch.}
    \label{tab:baseline_implementation}
\end{table}

\subsection{Baseline CRS Models} \label{apd:imp_crs}
We consider three baseline CRS models:
\begin{itemize}
\setlength{\itemsep}{0pt}
\setlength{\parsep}{0pt}
\setlength{\parskip}{0pt}
\item BARCOR \citep{wang2022barcor}: It Implements both the recommendation and conversation modules in a unified BART \citep{lewis2019bart}. 
\item UniCRS \citep{wang2022towards}: It implements both the recommendation and conversation modules in a unified DialoGPT \citep{zhang2020dialogpt} with knowledge-enhanced soft prompting. 
\item Llama2 \citep{touvron2023llama}: It preform recommendation by fine-tune Llama2 on the target dataset. We use Llama2-7b-chat-hf version.
\end{itemize}
These models are either the most representative CRS models or the most representative large language model.
For BARCOR, since it is not open-sourced, we follow \citet{wang2023rethinking} and \citet{wang2023improving} to implement it.
For Llama2, we use Lora \citep{hulora} with rank 4 to fine-tune it on downstream datasets.

\subsection{False Negative Detection Baseline} \label{apd:imp_fns}
Recently, as part of their contributions, \citet{wei2024llmrec} introduced a false negative detection approach in traditional recommender systems to first retrieve potential false negative samples with a supervised recommender, and then use an LLM to score the relevance between the items and the user preference.
These items with high relevance scores will be used as augmented labels to construct the augmented dataset.
After that, they use a one-stage training scheme to mix a certain proportion (refers to $\omega_3$ in the original paper) of labels from the original dataset and labels from the augmented dataset together to train the recommender.

Our implementation of this baseline is as follows. Firstly, we use the base recommender trained on the target dataset itself to retrieve items.
Then, we use the same LLM relevance scorer as our approach to score the relevance between the items and the dialogues.
These items with high relevance scores will be used as augmented labels to construct the augmented dataset.
After that, the same two-stage training scheme as our approach is used.

There are two differences between our implementation and the original implementation. Firstly, in the item retrieval stage, we use a different recommender from the original implementation.
Secondly, we use a different training scheme from the original implementation.
The reason why we use such an implementation is that. For the item retrieval stage, since \citet{wei2024llmrec} and us do not share the same task, we can not use the same recommender as \citet{wei2024llmrec} to retrieve items. So, we simply use the base recommender to implement the idea of considering both collaborative information and semantic information in the item retrieval stage.
For the training stage, the reason why we use the same two-stage training scheme as our approach instead of the original one-stage training scheme is twofold. Firstly, as shown in Table \ref{tab:baseline_implementation}, we find that our two-stage training scheme works better than the original one-stage training scheme.
Secondly, it helps us to fairly compare our item retrieval approach that disregards collaborative information with their item retrieval approach that regards collaborative information, which helps us gain a deeper understanding of how bias introduced by collaborative information could affect the final performance of the recommender.

\section{Case Study for Recommendation} \label{apd:case_study}

To gain a qualitative understanding of how our approach enhances the recommender's performance, we present several cases in Tables \ref{tab:case-study}, \ref{tab:case-study2}, and \ref{tab:case-study3}. As shown in the first lines of these cases, without augmented labels, the recommender struggles to suggest items that align semantically with the user's preferences. For instance, in the second case, when a user requests action and funny movies, the recommender, before data augmentation, suggests dramas instead, which are not semantically relevant to the user preference.
This leads to an unsatisfactory user experience.
after applying data augmentation, the recommender effectively suggests items that match the user's preferences more precisely. For example, in the second case, after data augmentation, the recommender successfully recommends several action comedy movies when the user requests action and funny movies, which are semantically relevant to the user preference.
This leads to satisfactory user experience.

To understand the underlying reasons, we examined similar dialogues between users and the recommender from the training set. As shown in the second lines of these cases, a significant number of semantically relevant items are added to the augmented training dataset. This improvement makes the recommender more likely to suggest these semantically relevant items.

Meanwhile, in these cases, the items that are accepted by the user in the test set are added into the augmented label set for the user that expresses a similar preference, which indicates that these items are highly likely to be false negative samples. Without data augmentation, these items are treated as negative samples, which directly reduces the probability for the recommender to recommend these items when encountering a user with a similar preference, leading to decreased user satisfaction.

\section{Case Study for Dialogue Generation} \label{apd:case_study_generation}
Following the modular design principle proposed by \citet{zhang2024recwizard} (see Section \ref{sec:dis}), we independently optimize the recommendation module. Below, we first discuss the benefits of this design principle and then demonstrate how our approach—which improves recommendation quality—can also enhance dialogue generation performance.

\paragraph{Advantages of modular design.} 
While prior work often treats the recommendation and generation modules as an unified system \citep{chen2019towards, zhou2020improving, wang2022barcor, wang2022towards}, \citet{zhang2024recwizard} argue that the two modules can be treated separately. We strongly endorse this principle, as pairing an existing recommendation module with a more powerful generation module (e.g., an LLM-based generator) can significantly improve how recommendations are presented to users.

To illustrate this, Table \ref{tab:table_gen1} compares responses generated by BARCOR as a unified system versus responses produced by combining BARCOR’s recommendation module with a strong LLM-based generator (denoted as BARCOR (modular), where we use Llama2-7b-chat-hf for generation, see Table \ref{tab:prompt_llmgen} for the prompt). The results show that the unified BARCOR system produces responses with limited informative explanations due to its weaker generation module. In contrast, when BARCOR’s recommendation module is paired with an LLM-based generator, the responses become more detailed and engaging, leading to a better user experience.


\paragraph{How our approach improve dialogue generation performance.}
To further demonstrate how our approach can improve dialogue generation, we conduct a case study using Llama2-7b-chat-hf as the generation module and compare responses generated from two different recommendation modules: (1) the original BARCOR recommender and (2) the BARCOR recommender enhanced by our approach (see Table \ref{tab:prompt_llmgen} for the prompt).

As shown in Table \ref{tab:table_gen2}, the original BARCOR recommender fails to align with user preferences, resulting in misleading responses. For example, it incorrectly describes Dunkirk (2017) as having a mix of action and humor, despite the film containing almost no comedic elements. Such inaccuracies can degrade user trust and satisfaction. In contrast, our improved recommender provides more relevant recommendations, leading to responses that are both factually correct and coherent.

This case study highlights how enhancing recommendation quality can indirectly improve dialogue generation. Future work could further explore synergies between different recommendation and generation modules.

\section{An Example Dialogue of the ReDial Dataset and the User Simulator} \label{apd:example}
To illustrate the difference between the ReDial dataset and the user simulator, we present a dialogue example from the ReDial dataset in Table\ref{tab:example_redial} and the user simulator in Table\ref{tab:example_ievalm}, respectively. The user in the ReDial dataset tends to mention more items and provide less detailed textual preference, while the user simulator tends to provide more detailed textual preference and mention fewer items.

\begin{table*}[t]
\centering
\small

\resizebox{\linewidth}{!}{\begin{tabular}{p{0.45\textwidth}|p{0.25\textwidth}|p{0.25\textwidth}}
\hline
\textbf{A Dialogue Context in the Test Set} & \textbf{BARCOR} & \textbf{BARCOR+ours} \\ \hline
\raggedright \textbf{User:} Hi I am looking for a movie like Super Troopers (2001) & 
\begin{tabular}[t]{@{}l@{}} 
1. Bad Boys (1995) \\ 
2. The Waterboy (1998) \\
3. Jerry Maguire (1996) \\
4. Click (2006) \\
5. American Pie (1999) \end{tabular} & 
\begin{tabular}[t]{@{}l@{}} 
1. Super Troopers 2 (2018) \\
2. 21 Jump Street (2012) \\
3. 22 Jump Street (2014) \\
\textcolor{blue}{4. Lethal Weapon (1987)} \\
\textcolor{blue}{5. Police Academy (1984)} \end{tabular} \\ \hline
\end{tabular}}

\resizebox{\linewidth}{!}{\begin{tabular}{p{0.45\textwidth}|p{0.25\textwidth}|p{0.25\textwidth}}
\hline
\textbf{A Similar Example in the Training Set} & \textbf{Original Label} & \textbf{Augmented Labels} \\ \hline
\raggedright \textbf{User:} Hi there! \newline \textbf{Recommender:} hello, what kind of movies are you looking to watch? \newline \textbf{User:} Could you recommend a comedy movie? I'm looking for some silly cop movies for my husband, something like Super Troopers (2001) or Hot Fuzz (2007) & 
\begin{tabular}[t]{@{}l@{}} 
Lethal Weapon (1987) \\ 
  \\
  \\
  \\
  \\
  \end{tabular} & 
\begin{tabular}[t]{@{}l@{}} 
Let's Be Cops (2014) \\
Super Troopers 2 (2018) \\
21 Jump Street (2012) \\
\textcolor{blue}{Police Academy (1984)} \\
The Other Guys (2010) \\
... (other 8 movies)\\ \end{tabular} \\ \hline
\end{tabular}}

\caption{A case study. Lethal Weapon (1987) and Police Academy (1984) are finally accepted by the user.}
\label{tab:case-study}
\end{table*}

\begin{table*}[t]
\centering
\small

\resizebox{\linewidth}{!}{\begin{tabular}{p{0.45\textwidth}|p{0.25\textwidth}|p{0.25\textwidth}}
\hline
\textbf{A Dialogue Context in the Test Set} & \textbf{BARCOR} & \textbf{BARCOR+ours} \\ \hline
\raggedright \textbf{User:} Hello! How are you? \newline \textbf{Recommender:} Hi I am good How are you \newline \textbf{User:} I'm fine. Do you have any good suggestions for a good movie? \newline \textbf{Recommender:} What kinds of movies do you like? Do you like dramatic movies? \newline \textbf{User:} I love action and funny movies, so I looking for something that combines the two, & 
\begin{tabular}[t]{@{}l@{}} 
1. Titanic (1997) \\ 
2. Dunkirk (2017) \\
3. 50 First Dates (2004) \\
4. Juno (2007) \\
5. Mr. \& Mrs. Smith (2005) \end{tabular} & 
\begin{tabular}[t]{@{}l@{}} 
1. Jumanji (2017) \\
2. \textcolor{blue}{Rush Hour (1998)} \\
3. \textcolor{blue}{Deadpool (2016)} \\
4. Thor: Ragnarok (2017) \\
5. Black Panther (2018) \end{tabular} \\ \hline
\end{tabular}}

\resizebox{\linewidth}{!}{\begin{tabular}{p{0.45\textwidth}|p{0.25\textwidth}|p{0.25\textwidth}}
\hline
\textbf{A Similar Example in the Training Set} & \textbf{Original Label} & \textbf{Augmented Labels} \\ \hline
\raggedright \textbf{User:} Hello \newline \textbf{Recommender:} Hello! how are you? \newline \textbf{User:} I'm good thanks \newline \textbf{Recommender:} What can I recommend for you today? \newline \textbf{User:} I enjoy action and comedy movies. & 
\begin{tabular}[t]{@{}l@{}} 
Superbad (2007) \\ 
  \\
  \\
  \\
  \\
  \end{tabular} & 
\begin{tabular}[t]{@{}l@{}} 
Spy (2015) \\
Knight and Day (2010) \\
Central Intelligence (2016) \\
\textcolor{blue}{Rush Hour (1998)} \\
Let's Be Cops (2014) \\
... (other 27 movies)\\ \end{tabular} \\ \hline
\end{tabular}}

\caption{A case study. Rush Hour (1987) and Deadpool (2016) are finally accepted by the user.}
\label{tab:case-study2}
\end{table*}

\begin{table*}[t]
\centering
\small

\resizebox{\linewidth}{!}{\begin{tabular}{p{0.45\textwidth}|p{0.25\textwidth}|p{0.25\textwidth}}
\hline
\textbf{A Dialogue Context in the Test Set} & \textbf{BARCOR} & \textbf{BARCOR+ours} \\ \hline
\raggedright \textbf{Recommender:} Hello! Do you like comedies? \newline \textbf{User:} I do! I really enjoy stoner comedies! \newline \textbf{Recommender:} Me too! \newline \textbf{User:} Know any good ones? & 
\begin{tabular}[t]{@{}l@{}} 
1. Dumb and Dumber (1994) \\ 
2. Bridesmaids (2011) \\
3. Superbad (2007) \\
4. The Heat (2013) \\
5. Office Space (1999) \end{tabular} & 
\begin{tabular}[t]{@{}l@{}} 
1. The Big Lebowski \\
2. This Is the End (2013) \\
3. Dazed and Confused (1993) \\
4. \textcolor{blue}{Pineapple Express (2008)} \\
5. Superbad (2007) \end{tabular} \\ \hline
\end{tabular}}

\resizebox{\linewidth}{!}{\begin{tabular}{p{0.45\textwidth}|p{0.25\textwidth}|p{0.25\textwidth}}
\hline
\textbf{A Similar Example in the Training Set} & \textbf{Original Label} & \textbf{Augmented Labels} \\ \hline
\raggedright \textbf{User:} Hey! hows the day going \newline \textbf{Recommender:} Pretty good and yours/ \newline \textbf{User:} would you be able to recommend some good movies today? im pretty chill here \newline \textbf{Recommender:} Me too, do you want some stoner comedy suggestions? \newline \textbf{User:} Actually yes that would be lovely & 
\begin{tabular}[t]{@{}l@{}} 
Super Troopers (2001) \\ 
  \\
  \\
  \\
  \\
  \end{tabular} & 
\begin{tabular}[t]{@{}l@{}} 
Up in Smoke (1978) \\
Harold \& Kumar \\
How High (2001) \\
\textcolor{blue}{Pineapple Express (2008)} \\
The Big Lebowski \\
... (other 27 movies)\\ \end{tabular} \\ \hline
\end{tabular}}

\caption{A case study. Pineapple Express (2008) is finally accepted by the user.}
\label{tab:case-study3}
\end{table*}

\begin{table*}[t]
    \centering
    \small
    \begin{tabular}{c|p{13cm}}
    \hline
    Context & User: Hi I am looking for a movie like Super Troopers (2001), please give me a suggestion. \\
    \hline
    \multirow{2}{*}{Responses} &\textbf{BARCOR:} Hi, I have some good movies. Have you seen 21 Jump Street (2012)? \\
    \cline{2-2}
    & \textbf{BARCOR (modular):} Sure, I'd be happy to help! Based on your dialogue history, it seems like you're looking for a movie similar to Super Troopers (2001). Given that, I would recommend the movie 21 Jump Street (2012) to you. Here's why: 21 Jump Street is a comedy film that, like Super Troopers, is a spoof of action movies and law enforcement. It features a similar blend of humor, action, and memorable characters, making it a great match for fans of Super Troopers. The movie also has a similar tone and style to Super Troopers, with a focus on absurd humor and over-the-top situations. I hope you enjoy 21 Jump Street! Let me know if you have any other questions or if you'd like more recommendations.\\
    
    \hline

    \end{tabular}
    \caption{A case study of generated response. BARCOR denotes treating BARCOR as a unified system, while BARCOR (modular) denotes combining BARCOR recommender with the LLM generator.}
    \label{tab:table_gen1}
\end{table*}

\begin{table*}[t]
    \centering
    \small
    \begin{tabular}{c|p{13cm}}
    \hline
    Context & User: Hello! How are you? 
    \newline Recommender: Hi I am good How are you 
    \newline User: I’m fine. Do you have any good suggestions for a good movie? 
    \newline Recommender: What kinds of movies do you like? Do you like dramatic movies? 
    \newline User: I love action and funny movies, so I looking for something that combines the two, 
    \\
    \hline
    \multirow{2}{*}{Responses} &\textbf{BARCOR:} Great! Based on your interest in action and funny movies, I would recommend "Titanic" (1997) and "Dunkirk" (2017). Both of these movies have a mix of action and humor that you might enjoy. "Titanic" is a classic romantic drama with plenty of action-packed scenes, while "Dunkirk" is a historical drama that also has some thrilling moments. I think you'll find them both to be great choices!
 \\
    \cline{2-2}
    & \textbf{BARCOR+ours:} Great! Based on your interest in action and funny movies, I would recommend "Jumanji" (2017) and "Rush Hour" (1998). Both of these movies have a great balance of action and humor that I think you'll enjoy. "Jumanji" is a thrilling adventure film with plenty of laughs, while "Rush Hour" is a classic buddy cop movie with a great mix of action and comedy. I hope you check them out and let me know what you think! \\
    
    \hline

    \end{tabular}
    \caption{A case study of generated response. BARCOR+ours denotes applying our approach on BARCOR recommender.}
    \label{tab:table_gen2}
\end{table*}

\begin{table*}[t]
    \centering
    \small
    \begin{tabular}{l|p{12.8cm}}
        \hline
        \textbf{User:}&Hello\\
        \hline
        \textbf{Recommender}:&hi\\
        \hline
        \textbf{User:}&i'm open to any movie What would you recommend me? i liked \textcolor{blue}{Avengers: Infinity War (2018)}\\
        \hline
        \textbf{Recommender:}&The Big Lebowski is my fav\\
        \hline
        \textbf{User:}&i also like scary movies i watched \textcolor{blue}{The Big Lebowski}\\
        \hline
        \textbf{Recommender:}&how about House of 1000 Corpses (2003)\\
        \hline
        \textbf{User:}&i also watched \textcolor{blue}{It (2017)} NO, i have not seen that one i guess i would try that one tomorrow i also watched \textcolor{blue}{It (1990)} the old one Which another one would you recommend\\
        \hline
        \textbf{Recommender:}& its by rob Zombie \\
        \hline
        \textbf{User:}&A funny one Like \textcolor{blue}{Click (2006)} by Adan Sandler \\
        \hline
    \end{tabular}
    \caption{An example from the ReDial dataset. Multiple items are mentioned by the user.}
    \label{tab:example_redial}
\end{table*}

\begin{table*}[t]
    \centering
    \small
    \begin{tabular}{l|p{12.8cm}}
        \hline
        \textbf{User:}&Hi I am looking for a movie like Super Troopers (2001).\\
        \hline
        \textbf{Recommender}:&Hi, I have some good movies.\\
        \hline
        \textbf{User:}&Great! I love movies with a similar comedic style and maybe some wacky characters. What do you have in mind?\\
        \hline
        \textbf{Recommender:}&Have you seen Bridesmaids (2011)?\\
        \hline
        \textbf{User:}&No, I haven't seen that one. I'm really looking for something with a more slapstick, over-the-top kind of humor, maybe with a focus on a group of misfits trying to succeed in a challenging environment.\\
        \hline
        \textbf{Recommender:}&Have you seen Superbad (2007)?\\
        \hline
        \textbf{User:}&Sorry, No, I haven't seen that one either. I'm looking for something with a bit more of a classic, buddy-cop feel,  with lots of physical comedy and maybe some training montage scenes.\\
        \hline
        \textbf{Recommender:}&Rush Hour (1998) is a good one.\\
        \hline
        \textbf{User:}&Sorry,  I'm looking for something a bit more lighthearted and silly, with a focus on a group of recruits learning the ropes in a somewhat unconventional way.\\
        \hline
    \end{tabular}
    \caption{An example from the user simulator. The user simulator provide detailed textual preference.}
    \label{tab:example_ievalm}
\end{table*}

\begin{table*}
    \centering
    \begin{tabular}{p{14cm}}
        \hline
        Given dialogue history between a user and a recommender, generate a response that recommend the target items for the user. \newline $\sim\sim\sim\sim\sim\sim\sim\sim\sim\sim\sim\sim\sim\sim\sim\sim$ \newline dialogue history: \newline \textbf{\{context\} }\newline$\sim\sim\sim\sim\sim\sim\sim\sim\sim\sim\sim\sim\sim\sim\sim\sim$ \newline items to recommend: \newline \textbf{\{item\}}\\
        \hline
    \end{tabular}
    \caption{Prompt template for the LLM generator}
    \label{tab:prompt_llmgen}
\end{table*}


\begin{table*}
    \centering
    \begin{tabular}{p{14cm}}
        \hline
You are provided with a dialogue history between a user and a recommender system, along with the system's top k movie recommendations. Your task is to score each recommendation on a scale from 0 to 4 based on how well it aligns with the user's expressed preferences. \newline $\sim\sim\sim\sim\sim\sim\sim$ \newline 4: The recommendation perfectly aligns with all expressed aspects of the user's preferences.\newline3: The recommendation aligns well with the main aspects of the user's preferences but lacks one or two minor aspects.\newline2: The recommendation aligns with some but not all major aspects of the user's preferences.\newline1: The recommendation aligns minimally with the user's preferences, missing several key aspects.\newline0: The recommendation does not align with the user's preferences at all. \newline You can give non-integer scores if necessary \newline $\sim\sim\sim\sim\sim\sim\sim$ \newline Here is an illustration of the scoring criteria.: \newline 
If a user said, `I like movies similar to Super Troopers (2001), especially its humorous style!' that may indicate the user's preference for the comedy genre, absurd humor, and perhaps law enforcement themes. Therefore, if a recommender system recommends movies such as `Hot Fuzz (2007)' or `The Other Guys (2010),' those should be 4-point recommendations because they align with the user's preferences for the comedy genre, absurd humor styles, and law enforcement themes. If a recommender system recommends a movie like `Superbad (2007),' this should be a 3-point recommendation since it aligns with the user's preference for the comedy genre and absurd humor styles but lacks a law enforcement theme. If a recommender system recommends a movie like `Police Story (1985),' this may be a 2-point recommendation since it has law enforcement themes and some comedy elements due to Jackie Chan's acting style. If a recommender system recommends a movie like `Infernal Affairs (2002),' this may be a 1-point recommendation since it has law enforcement themes but lacks any comedy or humor elements. If a recommender system recommends a movie like `It (2017),' this should be a 0-point recommendation since `It (2017)' is not relevant to the user's preferences. \newline Dialogue history: \newline \textbf{\{context\} }
\newline recommender's recommendation: \newline \textbf{\{rec\}} \newline Give brief reason and end with a JSON format as follows: \newline \{"<movie>": <score>, "<movie>": <score>, ...\} \newline Note: Replace <movie> and <score> with the movie name (year if exist) and actual score you have assigned to each movie.\\
        \hline
    \end{tabular}
    \caption{Prompt template for relevance estimation}
    \label{tab:prompt_llmreceval}
\end{table*}

\end{document}